\documentclass{article}
\usepackage{amsfonts}
\usepackage{amsmath}
\usepackage{graphicx}

\newtheorem{theorem}{Theorem}
\newtheorem{acknowledgement}{Acknowledgement}
\newtheorem{definition}{Definition}
\newtheorem{remark}{Remark}

\begin{document}

\title{On the electrical current distributions for the generalized Ohm's Law}
\author{M. P. Ramirez T. \\
Facultad de Ingenieria de la Universidad La Salle,\\
Benjamin Franklin 47, Col. Condesa, C.P. 06140, Mexico.\\
marco.ramirez@lasallistas.org.mx}
\maketitle

\begin{abstract}
The paper studies a particular class of analytic solutions for the
Generalized Ohm's Law, approached by means of the so called formal powers of
the Pseudoanalytic Function Theory. The reader will find a description of
the electrical current distributions inside bounded domains, within
inhomogeneous media, and their corresponding electric potentials near the
boundary. Finally, it is described a technique for approaching
separable-variables conductivity functions, a requisite when applying the
constructive methods posed in this work.

\end{abstract}

\section{Introduction}

The study of the generalized Ohm's Law

\begin{equation}
\text{div}\left( \sigma \text{grad}u\right) =0,  \label{int00}
\end{equation}%
where $\sigma $ denotes the electrical conductivity function and $u$ is the
electric potential, is the base for well understanding a wide class of
problems in Electromagnetic Theory, just as the Electrical Impedance
Tomography, the name given in medical imaging for the inverse problem posed
by Calderon \cite{calderon} in 1980. Yet, for many decades, the mathematical
complexity of (\ref{int00}) imposed so difficult challenges to the
researchers, that the structure of its general solution in analytic form
remained unknown. It was until 2006 that K. Astala and L. P\"{a}iv\"{a}rinta 
\cite{astala} discovered that the two-dimensional case of (\ref{int00}) was
closely related with a Vekua equation \cite{vekua}, and in 2007 V.
Kravchenko et al. \cite{kr-o}, based upon elements of Pseudoanalytic
Function Theory \cite{bers}, achieved to pose what could be considered the
first general solution in analytic form of (\ref{int00}) for the plain, when
the conductivity function $\sigma $ belongs to a special class of functions.

Virtually, these two discoveries were the departure point for developing a
completely new theory for the generalized Ohm's Law, mainly because they
allowed to research a wide sort of electromagnetic phenomena that had
remained out of range for the mathematical tools known previously.

In this paper we discuss the possibility of considering the \emph{formal
powers }\cite{bers}, as a new scope for analyzing the electrical current
distributions inside inhomogeneous media in bounded domains, since it is
bias their linear combination that we can approach the general solution for
the two-dimensional case of (\ref{int00}) \cite{iaeng09}, and they are also
useful for constructing an infinite set of analytic solutions for its
three-dimensional case \cite{wce10}.

Starting with the elements of Pseudoanalytic Function Theory, and of
Quaternionic Analysis, we briefly expose an idea for rewriting the
three-dimensional case of (\ref{int00}) in a quaternionic equation, in order
to pose the structure of its general solution by means of a generalization
of the Bers generating pair, in Complex Analysis.

Eventually, we focus our attention on the plane by considering one example
in which a spacial variable is fixed, and the conductivity $\sigma $ adopts
an exponential form. Then we trace the electrical current density patches,
obtained from the solutions of the Vekua equation in terms of Taylor series
in formal powers, and we show that, from an adequate point of view, the
observed patches might keep a sort of regular dynamics when flowing through
this inhomogeneous medium, once they are compared to those traced for an
homogeneous case.

The work closes with a basic idea for approaching separable-variables
conductivity functions, since this is the central requirement if we are to
apply the exposed mathematical ideas for approaching solutions of (\ref%
{int00}), and to analyze their meaning in terms of electrical current flows.

\section{Preliminaries}

\subsection{Elements of pseudoanalytic functions}

Following L. Bers \cite{bers}, a pair of complex-valued functions $F=\text{Re%
}\left( F\right) +i\text{Im}\left( F\right) $ and $G=\text{Re}\left(
G\right) +i\text{Im}\left( G\right) $ will be called a \emph{generating pair 
}if the following condition holds:%
\begin{equation}
\text{Im}\left( \overline{F}G\right) >0.  \label{pre03}
\end{equation}%
Here $i$ denotes the standard imaginary unit $i^{2}=-1$, whereas $\overline{F%
}$ represents the complex conjugation of $F:\overline{F}=\text{Re}\left(
F\right) -i\text{Im}\left( F\right) .$ Thus, any complex-valued function $W$
can be represented as the linear combination of the generating pair $\left(
F,G\right) $:%
\begin{equation}
W=\phi F+\psi G,  \label{pre04}
\end{equation}%
where $\phi $ and $\psi $ are both real-valued functions. Hence the \emph{%
derivative in the sense of Bers, }or $\left( F,G\right) $\emph{-derivative},
of a complex-valued function $W$ will be defined as%
\begin{equation}
\partial _{\left( F,G\right) }W=\left( \partial _{z}\phi \right) F+\left(
\partial _{z}\psi \right) G,  \label{pre05}
\end{equation}%
where $\partial _{z}=\frac{\partial }{\partial x}-i\frac{\partial }{\partial
y}$, and it will exist if and only if%
\begin{equation}
\left( \partial _{\overline{z}}\phi \right) F+\left( \partial _{\overline{z}%
}\psi \right) G=0,  \label{pre06}
\end{equation}%
where $\partial _{\overline{z}}=\frac{\partial }{\partial x}+i\frac{\partial 
}{\partial y}.$ Notice even the operators $\partial _{z}$ and $\partial _{%
\overline{z}}$ are usually defined including the coefficient $\frac{1}{2},$
it will result somehow more convenient for this paper to work without it.

By introducing the notations%
\begin{gather}
A_{\left( F,G\right) }=-\frac{\overline{F}\partial _{z}G-\overline{G}%
\partial _{z}F}{F\overline{G}-\overline{F}G},\text{ \ \ }a_{\left(
F,G\right) }=-\frac{\overline{F}\partial _{\overline{z}}G-\overline{G}%
\partial _{\overline{z}}F}{F\overline{G}-\overline{F}G},  \label{pre07} \\
B_{\left( F,G\right) }=\frac{F\partial _{z}G-G\partial _{z}F}{F\overline{G}-%
\overline{F}G},\text{ \ \ }b_{\left( F,G\right) }=\frac{F\partial _{%
\overline{z}}G-G\partial _{\overline{z}}F}{F\overline{G}-\overline{F}G}; 
\notag
\end{gather}%
the $\left( F,G\right) $-derivative of $W$ (\ref{pre05}) can be written as%
\begin{equation}
\partial _{\left( F,G\right) }W=A_{\left( F,G\right) }W+B_{\left( F,G\right)
}\overline{W},  \label{pre08}
\end{equation}%
and the condition (\ref{pre06}) will turn into%
\begin{equation}
\partial _{\overline{z}}W-a_{\left( F,G\right) }W-b_{\left( F,G\right) }%
\overline{W}=0.  \label{pre09}
\end{equation}

The last differential equation is known as the \emph{Vekua equation }\cite%
{vekua}, and it will play a very important roll in our further discussions.
The functions defined in (\ref{pre07}) are known as the \emph{characteristic
coefficients }of the generating pair $\left( F,G\right) $, and every
complex-valued function $W$ satisfying (\ref{pre09}) will be referred as an $%
\left( F,G\right) $\emph{-pseudoanalytic} \emph{function}.

The following statements were originally posed in \cite{bers}. Another
authors will be cited explicitly.

\begin{remark}
The functions conforming the generating pair $\left( F,G\right) $ are $%
\left( F,G\right) $-pseudoanalytic, and their $\left( F,G\right) $%
-derivatives are $\partial _{\left( F,G\right) }F=\partial _{\left(
F,G\right) }G=0.$
\end{remark}

\begin{remark}
Let $p$ be a non-vanishing function inside some domain $\Omega $. The pair
of functions $F=p$ and $G=\frac{i}{p}$ satisfy the condition (\ref{pre03}),
so they constitute a Bers generating pair, and their characteristic
coefficients are%
\begin{gather}
A_{\left( F,G\right) }=a_{\left( F,G\right) }=0,  \label{pre10} \\
B_{\left( F,G\right) }=\frac{\partial _{z}p}{p},\text{ \ \ }b_{\left(
F,G\right) }=\frac{\partial _{\overline{z}}p}{p}.  \notag
\end{gather}
\end{remark}

\begin{theorem}
\cite{kpa} Let $p$ be a non-vanishing function within some domain $\Omega $.
The pair of real-valued functions $\phi $ and $\psi $ will be solutions of
the system%
\begin{equation}
\frac{\partial }{\partial x}\phi =\frac{1}{p^{2}}\frac{\partial }{\partial y}%
\psi ,\text{ \ \ }\frac{\partial }{\partial y}\phi =-\frac{1}{p^{2}}\frac{%
\partial }{\partial x}\psi ,  \label{pre19a}
\end{equation}%
if and only if $W=p\phi +\frac{i}{p}\psi $ is solution of the Vekua equation%
\begin{equation*}
\partial _{\overline{z}}W-\frac{\partial _{\overline{z}}p}{p}\overline{W}=0.
\end{equation*}%
A pair of functions $\phi $ and $\psi $ satisfying (\ref{pre19a}) is called
a $p$\emph{-analytic system \cite{polo}.}
\end{theorem}

\begin{definition}
Let $\left( F_{0},G_{0}\right) $ and $\left( F_{1},G_{1}\right) $ be two
generating pairs, and let their characteristic coefficients satisfy%
\begin{equation}
a_{\left( F_{0},G_{0}\right) }=a_{\left( F_{1},G_{1}\right) }\text{ \ \ \
and \ \ \ }B_{(F_{0},G_{0})}=-b_{(G_{1},F_{1})}.  \label{pre11}
\end{equation}%
The generating pair $\left( F_{1},G_{1}\right) $ will be called a \emph{%
successor pair }of $\left( F_{0},G_{0}\right) ,$ as well as the pair $\left(
F_{0},G_{0}\right) $ will be called a \emph{predecessor }of $\left(
F_{1},G_{1}\right) $.
\end{definition}

\begin{theorem}
Let the complex-valued function $W$ be $\left( F_{0},G_{0}\right) $%
-pseudoanalytic, and let the generating pair $\left( F_{1},G_{1}\right) $ be
a successor pair of $\left( F_{0},G_{0}\right) $. Hence, the $\left(
F_{0},G_{0}\right) $-derivative of $W$ will be an $\left( F_{1},G_{1}\right) 
$-pseudoanalytic function.
\end{theorem}

\begin{definition}
Let the pairs of functions belonging to the set 
\begin{equation}
\left\{ \left( F_{n},G_{n}\right) :n=0,\pm 1,\pm 2,...\right\}  \label{pre12}
\end{equation}%
be all generating pairs, and let every $\left( F_{n+1},G_{n+1}\right) $ be a
successor pair of $\left( F_{n},G_{n}\right) $. Then the set (\ref{pre12})
is called a \emph{generating sequence}. If the generating pair $\left(
F,G\right) =\left( F_{0},G_{0}\right) $, we say that $\left( F,G\right) $ is 
\emph{embedded }in the generating sequence (\ref{pre12}).
\end{definition}

\begin{definition}
The generating pairs $\left( F_{n},G_{n}\right) $ and $\left( F_{n}^{\prime
},G_{n}^{\prime }\right) $ are called \emph{equivalent} if they posses the
same characteristic coefficients (\ref{pre07}).
\end{definition}

\begin{definition}
A generating sequence $\left\{ \left( F_{n},G_{n}\right) \right\} $ is
called \emph{periodic,} with period $\alpha >0,$ if the generating pairs $%
\left( F_{n},G_{n}\right) $ and $\left( F_{n+\alpha },G_{n+\alpha }\right) $
are equivalent.
\end{definition}

\begin{definition}
Let the complex-valued function $W$ be $\left( F,G\right) $-pseudoanalytic,
and let $\left\{ \left( F_{n},G_{n}\right) \right\} $ be a generating
sequence in which the generating pair $\left( F,G\right) $ is embedded. The
higher derivatives in the sense of Bers of $W$ will be expressed as%
\begin{gather*}
W^{[0]}=W; \\
W^{[n+1]}=\partial _{_{\left( F_{n},G_{n}\right) }}W^{[n]};\text{ }%
n=0,1,2,...
\end{gather*}
\end{definition}

L. Bers also introduced the notion of the $\left( F,G\right) $-integral for
a complex-valued function $W$. The following statements describe its
structure, the necessary conditions for its existence and some of its
properties.

\begin{definition}
Let $\left( F_{0},G_{0}\right) $ be a generating pair. Its \emph{adjoint
pair }$\left( F_{0}^{\ast },G_{0}^{\ast }\right) $ will be defined according
to the formulas%
\begin{equation*}
F_{0}^{\ast }=-\frac{2\overline{F}_{0}}{F_{0}\overline{G}_{0}-\overline{F}%
_{0}G_{0}},\ \ G_{0}^{\ast }=\frac{2\overline{G}_{0}}{F_{0}\overline{G}_{0}-%
\overline{F}_{0}G_{0}}.
\end{equation*}%
In particular, if $p$ is a non-vanishing function inside a domain $\Omega $,
the adjoint pair of $F_{0}=p$, $G_{0}=\frac{i}{p}$ will have the form%
\begin{equation*}
F_{0}^{\ast }=-ip,\text{ \ \ }G_{0}^{\ast }=p^{-1}.
\end{equation*}
\end{definition}

\begin{definition}
The $\left( F_{0},G_{0}\right) $\emph{-integral} of a complex-valued
function $W$ is defined as%
\begin{gather*}
\int_{\Lambda }Wd_{\left( F_{0},G_{0}\right) }z=G_{0} \text{Re} \int_{\Lambda
}F_{0}^{\ast }Wdz+F_{0} \text{Re} \int_{\Lambda }G_{0}^{\ast }Wdz,
\end{gather*}%
where $\Lambda $ is any rectifiable curve inside some domain $\Omega $,
going from $z_{0}$ until $z,$ and it will exist iff $W$ satisfies%
\begin{equation*}
\text{Re}\oint G_{0}^{\ast }Wdz+i\text{Re}\oint F_{0}^{\ast }Wdz=0.
\end{equation*}
\end{definition}

\begin{theorem}
The $\left( F_{0},G_{0}\right) $-derivative of an $\left( F_{0},G_{0}\right) 
$-pseudoanalytic function $W$ will be $\left( F_{0},G_{0}\right) $%
-integrable.
\end{theorem}

\begin{remark}
If $W=\phi F_{n}+\psi G_{n}$ is an $\left( F_{n},G_{n}\right) $%
-pseudoanalytic function inside $\Omega $, its $\left( F_{n},G_{n}\right) $%
-derivative will be $\left( F_{n},G_{n}\right) $-integrable. Indeed%
\begin{equation*}
\int_{z_{0}}^{z}\partial _{\left( F_{n},G_{n}\right) }W\left( z\right)
d_{\left( F_{n},G_{n}\right) }z=W\left( z\right) -\phi \left( z_{0}\right)
F_{n}\left( z\right) -\psi \left( z_{0}\right) G_{n}\left( z\right) ,
\end{equation*}%
where $z_{0}$ is a fixed point. Moreover, by Remark 1 the $\left(
F_{n},G_{n}\right) $-derivatives of $F_{n}$ and $G_{n}$ vanish identically,
thus the last integral expression represents the $\left( F_{n},G_{n}\right) $%
-\emph{antiderivative }of $\partial _{\left( F_{n},G_{n}\right) }W\left(
z\right) .$
\end{remark}

The following paragraphs will expose some definitions and properties of the
so called \emph{formal powers}, whose physical implications will provide
most material for this work.

\begin{definition}
The formal power $Z_{n}^{\left( 0\right) }\left( a_{0},z_{0};z\right) $
belonging to the generating pair $\left( F_{n},G_{n}\right) $, with exponent 
$0$, complex coefficient $a_{0}$, center at the fixed point $z_{0}$ and
depending upon $z=x+iy$, is defined by the linear combination of $F_{n}$ and 
$G_{n}$ according to the expression%
\begin{equation*}
Z_{n}^{\left( 0\right) }\left( a_{0},z_{0};z\right) =\lambda F_{n}+\mu G_{n},
\end{equation*}%
where $\lambda $ and $\mu $ are real constants such that%
\begin{equation*}
\lambda F_{n}\left( z_{0}\right) +\mu G_{n}\left( z_{0}\right) =a_{0}.
\end{equation*}%
The formal powers with higher exponents are defined by the recursive formulas%
\begin{equation*}
Z_{n}^{\left( m+1\right) }\left( a_{m},z_{0};z\right) =\left( m+1\right)
\int_{\Lambda }Z_{n+1}^{\left( m\right) }\left( a_{m},z_{0};z\right)
d_{\left( F_{n},G_{n}\right) }z;
\end{equation*}%
where $m,n=0,1,2,...$
\end{definition}

\begin{theorem}
The formal powers posses the following properties:\newline
\textbf{1)} $Z_{n}^{\left( m\right) }\left( a_{m},z_{0};z\right) $ is $%
\left( F_{n},G_{n}\right) $-pseudoanalytic.\newline
\textbf{2)} If $a^{\prime }$ and $a^{\prime \prime }$ are real constants,
then%
\begin{equation*}
Z_{n}^{\left( m\right) }\left( a^{\prime }+ia^{\prime \prime
},z_{0};z\right) =a^{\prime }Z_{n}^{\left( m\right) }\left( 1,z_{0};z\right)
+a^{\prime \prime }Z_{n}^{\left( m\right) }\left( i,z_{0};z\right) ,
\end{equation*}%
for $m,n=0,1,2,...$\newline
\textbf{3)} When $z\rightarrow z_{0}$ we have that 
\begin{equation*}
Z_{n}^{\left( m\right) }\left( a_{m},z_{0};z\right) \rightarrow a_{m}\left(
z-z_{0}\right) ^{m}.
\end{equation*}
\end{theorem}

\begin{theorem}
Let $W$ be an $\left( F,G\right) $-pseudoanalytic function. Then, it accepts
the expansion%
\begin{equation}
W=\sum_{m=0}^{\infty }Z^{\left( m\right) }\left( a_{m},z_{0};z\right) ,
\label{pre13}
\end{equation}%
where the absence of the subindex "$n$" in the formal powers $Z^{\left(
m\right) }\left( a_{m},z_{0};z\right) $ implies that all formal powers
belong to the same generating pair. The coefficients $a_{m}$ will be given
by the formulas%
\begin{equation*}
a_{m}=\frac{W^{\left[ m\right] }\left( z_{0}\right) }{n!}.
\end{equation*}%
The expansion (\ref{pre13}) is called \emph{Taylor series in formal powers }%
of $W$.
\end{theorem}

\begin{remark}
Since every $\left( F,G\right) $-pseudoanalytic function $W$ can be
expressed by means of the Taylor series (\ref{pre13}), this expansion is in
fact an analytical representation for the general solution of the Vekua
equation (\ref{pre09}), being the formal powers%
\begin{equation*}
Z^{\left( m\right) }\left( 1,z_{0};z\right) \text{ \ \ and \ \ }Z^{\left(
m\right) }\left( i,z_{0};z\right) ,
\end{equation*}%
by virtue of Theorem 15, number (\textbf{2)}, a base for the set of its
solutions.
\end{remark}

\subsubsection{An alternative path for introducing the concept of Bers
generating pair}

Following \cite{k-r-b}, let $\phi $ be a real-valued function and let $F$ be
a complex-valued function. Consider the equality 
\begin{equation}
\left( \partial _{\overline{z}}-a-b\mathbf{C}\right) \left( \phi F\right)
=\left( \partial _{\overline{z}}\phi \right) F;  \label{pre14}
\end{equation}%
where $a$ and $b$ are complex-valued functions, and $\mathbf{C}$ denotes the
complex conjugation operator acting upon $F$ as $\mathbf{C}F=\overline{F}$.
The partial differential operator in the left side of the equation is
clearly the one corresponding to a Vekua equation. A simple calculation will
show that (\ref{pre14}) will be valid if and only if the complex-valued
function $F$ is a particular solution of%
\begin{equation}
\partial _{\overline{z}}F-aF-b\overline{F}=0.  \label{pre15}
\end{equation}

In the same way, let $\psi $ be a real-valued function and let $G$ be a
particular solution of (\ref{pre15}). Thus the equality%
\begin{equation*}
\left( \partial _{\overline{z}}-a-b\mathbf{C}\right) \left( \psi G\right)
=\left( \partial _{\overline{z}}\psi \right) G
\end{equation*}%
will hold. By adding this equation with (\ref{pre14}), and introducing the
notation $W=\phi F+\psi G$, we will obtain%
\begin{equation*}
\partial _{\overline{z}}W-aW-b\overline{W}=\left( \partial _{\overline{z}%
}\phi \right) F+\left( \partial _{\overline{z}}\psi \right) G.
\end{equation*}%
Hence%
\begin{equation}
\partial _{\overline{z}}W-aW-b\overline{W}=0  \label{pre17}
\end{equation}%
if and only if%
\begin{equation*}
\left( \partial _{\overline{z}}\phi \right) F+\left( \partial _{\overline{z}%
}\psi \right) G=0.
\end{equation*}%
If we require the functions $F$ and $G$ to fulfil the condition (\ref{pre03}%
), we will arrive to the very definition of an $\left( F,G\right) $%
-pseudoanalytic function $W$. Indeed, additional calculations will show that
the functions $a$ and $b$ are precisely the characteristic coefficients $%
a_{\left( F,G\right) }$ and $b_{\left( F,G\right) }$ defined in (\ref{pre07}%
), thus (\ref{pre17}) coincides with (\ref{pre09}).

This alternative procedure will be useful for introducing the notion of 
\emph{Bers generating sets} for the solutions of the three-dimensional
quaternionic Generalized Ohm's Law.

\subsection{Elements of Quaternionic Analysis}

The algebra of real quaternions will be denoted by $\mathbb{H}\left( \mathbb{%
R}\right) $ (see e.g. \cite{kbook}). Every element belonging to this set
will have the form $q=\sum_{k=0}^{3}q_{k}\mathbf{e}_{k},$ where $%
q_{k}=q_{k}\left( x_{1},x_{2},x_{3}\right) ;$ $k=\overline{0,3}$ are in
general real-valued functions depending upon three spacial variables, $%
\mathbf{e}_{0}=1,$ and $\mathbf{e}_{k};$ $k=1,2,3$ are the standard
quaternionic units, possessing the following properties of multiplication:%
\begin{gather}
\mathbf{e}_{1}^{2}=\mathbf{e}_{2}^{2}=\mathbf{e}_{3}^{2}=-1,  \label{pre00}
\\
\mathbf{e}_{1}\mathbf{e}_{2}\mathbf{e}_{3}=-1.  \notag
\end{gather}%
It will be useful to introduce the auxiliary notation for an element $q\in 
\mathbb{H}\left( \mathbb{R}\right) $ 
\begin{equation*}
q=q_{0}+\overrightarrow{q},
\end{equation*}%
where clearly $\overrightarrow{q}=\sum_{k=0}^{3}q_{k}\mathbf{e}_{k}$. Thus $%
q_{0}$ will be named the \emph{scalar part }of the quaternion $q$, whereas $%
\overrightarrow{q}$ will be referred as the \emph{vectorial part} of $q.$ It
is important to point out that the set of \emph{purely vectorial }%
quaternionic functions such that $\left\{ q=\overrightarrow{q}:q\in \mathbb{H%
}\left( \mathbb{R}\right) \right\} $ conforms an isomorphism with the set of
three-dimensional Cartesian vectors $\mathbb{R}^{3}.$

As it can be easily inferred from (\ref{pre00}), the quaternionic product is
not commutative, hence the \emph{multiplication by the right hand-side} of
the quaternion $p$ by the quaternion $q$ will be written as%
\begin{equation*}
pq=M^{q}p.
\end{equation*}

\subsubsection{The Moisil-Theodoresco differential operator}

On the set of at least once-differentiable quaternionic-valued functions, it
is defined the Moisil-Theodoresco partial differential operator, that was
first introduced by Hamilton itself%
\begin{equation*}
D=\sum_{k=1}^{3}\mathbf{e}_{k}\frac{\partial }{\partial x_{k}}.
\end{equation*}%
By means of the isomorphism remarked before, the operator $D$ acts upon a
function $q\in \mathbb{H}\left( \mathbb{R}\right) $ according to the rule:%
\begin{equation}
Dq=\text{grad}q_{0}-\text{div}\overrightarrow{q}+\text{rot}%
\overrightarrow{q},  \label{pre02}
\end{equation}%
where "$\text{grad}$", "$\text{div}$" and "$\text{rot}$" are the
classical Cartesian operators, written using quaternionic notations (see
e.g. \cite{kbook}).

\subsubsection{Bers generating sets for the solutions of partial
differential equations}

Let us consider the quaternionic differential equation%
\begin{equation}
DQ+Qp=0,  \label{pre16}
\end{equation}%
where $p,Q\in \mathbb{H}\left( \mathbb{R}\right) .$ Employing the idea
exposed within the last paragraphs dedicated to the elements of
pseudoanalytic functions, let $Q$ be a particular solution of the equation (%
\ref{pre16}) and let $\varphi $ be a purely scalar function. A short
calculation will show that the following equation holds%
\begin{equation*}
D\left( \varphi Q\right) +\left( \varphi Q\right) p=\left( D\varphi \right)
Q.
\end{equation*}%
Hence, if we own a set of four linearly independent $Q_{k}\in \mathbb{H}%
\left( \mathbb{R}\right) ;~k=\overline{0,3}$, every one solution of (\ref%
{pre16}), we will be able to represent the general solution of (\ref{pre16})
by means of the linear combination of $Q_{k}$.

\begin{theorem}
\cite{k-r-b} Let $\left\{ Q_{k}\right\} _{k=0}^{3}\subset \mathbb{H}\left( 
\mathbb{R}\right) $ be a set of linearly independent solutions of the
equation (\ref{pre16}), thus its general solution can be written as%
\begin{equation*}
Q=\sum_{k=0}^{3}\varphi _{k}Q_{k},
\end{equation*}%
where $\left\{ \varphi _{k}\right\} _{k=0}^{3}$ $\subset \mathbb{R}$ are
scalar functions, all solutions of%
\begin{equation}
\sum_{k=0}^{3}\left( D\varphi _{k}\right) Q_{k}=0.  \label{pre18}
\end{equation}%
We shall call $\left\{ Q_{k}\right\} _{k=0}^{3}$ the \emph{Bers generating
set }for the solutions of (\ref{pre16}).
\end{theorem}

\section{Study of the generalized Ohm's Law for a separable-variables
conductivity function $\protect\sigma $}

As it was shown in \cite{wce10} and \cite{cce09}, introducing the notations%
\begin{equation}
\overrightarrow{\mathcal{E}}=-\sqrt{\sigma }\text{grad}u,\text{ }%
\overrightarrow{\mathbf{\sigma }}=\frac{\text{grad}\sqrt{\sigma }}{\sqrt{%
\sigma }};  \label{gol09}
\end{equation}%
the Ohm's Law (\ref{int00}) can be rewritten in quaternionic form:%
\begin{equation}
\left( D+M^{\overrightarrow{\mathbf{\sigma }}}\right) \overrightarrow{%
\mathcal{E}}=0.  \label{gol00}
\end{equation}%
Hence, by virtue of Theorem 18, its general solution will have the form%
\begin{equation}
\overrightarrow{\mathcal{E}}=\sum_{k=1}^{3}\varphi _{k}\overrightarrow{%
\mathcal{E}}_{k},  \label{gol01}
\end{equation}%
where $\left\{ \overrightarrow{\mathcal{E}}_{k}\right\} _{k=1}^{3}\subset 
\mathbb{H}\left( \mathbb{R}\right) $ is a set of linearly independent
solutions of (\ref{gol00}), and $\left\{ \mathcal{\varphi }_{k}\right\}
_{k=1}^{3}\mathbb{\subset R}$ are scalar functions, all solutions of%
\begin{equation}
\sum_{k=1}^{3}\left( D\varphi _{k}\right) \overrightarrow{\mathcal{E}}_{k}=0.
\label{gol02}
\end{equation}

Specifically, when the conductivity $\sigma $ is a separable-variables
function $\sigma =s_{1}\left( x_{1}\right) s_{2}\left( x_{2}\right)
s_{3}\left( x_{3}\right) $ the Bers generating set (\ref{gol01}) can be
constructed explicitly \cite{wce10} (it is worth of mention that a
separable-variables conductivity function, in polar coordinates, was
considered in the interesting work \cite{demi}, for studying the Generalized
Ohm's Law by means of different mathematical methods). For example, let $%
\overrightarrow{\mathcal{E}}_{1}=\mathbf{e}_{1}\mathcal{E}_{1},$ where $%
\mathcal{E}_{1}\in \mathbb{R}$. Substituting into (\ref{gol00}), we will
obtain the partial differential system%
\begin{equation*}
\frac{\partial }{\partial x_{1}}\mathcal{E}_{1}+s_{1}^{\prime }\mathcal{E}%
_{1}=0,\text{ }\frac{\partial }{\partial x_{2}}\mathcal{E}_{1}-s_{2}^{\prime
}\mathcal{E}_{1}=0,\text{ }\frac{\partial }{\partial x_{3}}\mathcal{E}%
_{1}-s_{3}^{\prime }\mathcal{E}_{1}=0,
\end{equation*}%
where $s_{k}^{\prime }=\frac{1}{s_{k}}\frac{\partial }{\partial x_{k}}%
s_{k};~k=1,2,3,$ for which a solution is%
\begin{equation*}
\mathcal{E}_{1}=e^{-\int s_{1}^{\prime }dx_{1}+\int s_{2}^{\prime
}dx_{2}+\int s_{3}^{\prime }dx_{3}}.
\end{equation*}%
Using the same idea, we will find out that the set of quaternionic-valued
functions%
\begin{eqnarray}
\overrightarrow{\mathcal{E}}_{1} &=&\mathbf{e}_{1}\mathcal{E}_{1}=\mathbf{e}%
_{1}e^{-\int s_{1}^{\prime }dx_{1}+\int s_{2}^{\prime }dx_{2}+\int
s_{3}^{\prime }dx_{3}},  \label{gol03} \\
\overrightarrow{\mathcal{E}}_{2} &=&\mathbf{e}_{2}\mathcal{E}_{2}=\mathbf{e}%
_{2}e^{\int s_{1}^{\prime }dx_{1}-\int s_{2}^{\prime }dx_{2}+\int
s_{3}^{\prime }dx_{3}},  \notag \\
\overrightarrow{\mathcal{E}}_{3} &=&\mathbf{e}_{3}\mathcal{E}_{3}=\mathbf{e}%
_{3}e^{\int s_{1}^{\prime }dx_{1}+\int s_{2}^{\prime }dx_{2}-\int
s_{3}^{\prime }dx_{3}},  \notag
\end{eqnarray}%
conforms a Bers generating set for the solutions of (\ref{gol00}).

We must find the general solution of (\ref{pre18}) if we are to pose the
general solution of (\ref{gol00}), but it is not clear yet how to achieve
this task. However, it is possible to construct an infinite set of solutions
for (\ref{gol00}) by means of the following procedure \cite{wce10}.

Suppose the scalar function $\varphi _{3}$ vanishes identically inside the
domain of interest $\Omega $. Then (\ref{pre18}) will turn into%
\begin{equation*}
\left( D\varphi _{1}\right) \overrightarrow{\mathcal{E}}_{1}+\left( D\varphi
_{2}\right) \overrightarrow{\mathcal{E}}_{2}=0,
\end{equation*}%
which will reach the system%
\begin{gather}
\frac{\partial }{\partial x_{1}}\varphi _{1}=-\frac{1}{p^{2}}\frac{\partial 
}{\partial x_{2}}\varphi _{2},\text{ \ \ }\frac{\partial }{\partial x_{2}}%
\varphi _{1}=\frac{1}{p^{2}}\frac{\partial }{\partial x_{1}}\varphi _{2},
\label{gol04} \\
\frac{\partial }{\partial x_{3}}\varphi _{1}=\frac{\partial }{\partial x_{3}}%
\varphi _{2}=0;  \notag
\end{gather}%
where $p=e^{-\int s_{1}^{\prime }dx_{1}+\int s_{2}^{\prime }dx_{2}}$. The
first pair of equations conforms precisely the $p$-analytic system \cite%
{polo} introduced in Theorem 3, thus its corresponding Vekua equation will
have the form%
\begin{equation}
\partial _{\overline{\zeta}}W-\frac{\partial_{\overline{\zeta}}p}{p}\overline{W%
}=0,  \label{gol10}
\end{equation}%
where $W=p\varphi _{1}+\frac{i}{p}\varphi _{2}$, and $\partial _{\overline{%
\zeta }}=\frac{\partial }{\partial x_{2}}+i\frac{\partial }{\partial x_{1}}.$

Moreover, based upon a result posed by L. Bers itself in \cite{bers}, and
latter generalized by V. Kravchenko in \cite{kpa}, slightly adapted for this
work, we are able to find in analytic form the generating sequence that will
allow us to build the formal powers for approaching the general solution of (%
\ref{gol10}) in terms of Taylor series.

\begin{theorem}
\cite{bers}\cite{kpa} Let $p$ be a non-vanishing separable-variables
function inside a domain $\Omega $, and let $F_{0}=p$, $G_{0}=\frac{i}{p}$.
Since $p=p_{1}\left( x_{1}\right) p_{2}\left( x_{2}\right) $ the generating
pair $\left( F_{0},G_{0}\right) $ is embedded within a periodic generating
sequence $\left\{ \left( F_{m},G_{m}\right) \right\} $, with period $2$,
such that%
\begin{equation*}
F_{m}=p_{1}\left( x_{1}\right) p_{2}\left( x_{2}\right) ,\text{ \ \ }G_{m}=%
\frac{i}{p_{1}\left( x_{1}\right) p_{2}\left( x_{2}\right) },
\end{equation*}%
when $m$ is an even number, and%
\begin{equation*}
F_{m}=\frac{p_{1}\left( x_{1}\right) }{p_{2}\left( x_{2}\right) },\text{ \ \ 
}G_{m}=i\frac{p_{2}\left( x_{2}\right) }{p_{1}\left( x_{1}\right) }
\end{equation*}%
when $m$ is odd.
\end{theorem}

It is evident that identical procedures can be employed for the cases when
the scalar function $\varphi _{1}$ vanishes identically in (\ref{pre18}),
and when $\varphi _{2}$ does; leading to an infinite set of solutions for
the three-dimensional quaternionic Generalized Ohm's Law (\ref{gol00}).

\subsection{Electrical current distributions in analytic form}

In order to pose an example for analyzing the behavior of the electrical
current distributions, obtained bias the formal powers method; let us
consider a conductivity function $\sigma $ with the form%
\begin{equation*}
\sigma =s_{1}\left( x_{1}\right) s_{2}\left( x_{2}\right) s_{3}\left(
x_{3}\right) =e^{2\sigma _{1}x_{1}+2\sigma _{2}x_{2}+2\sigma _{3}x_{3}},
\end{equation*}%
where $\sigma _{1},\sigma _{2}$ and $\sigma _{3}$ are all real constants,
and let $\varphi _{3}$ vanish identically inside $\Omega .$ In concordance
with (\ref{gol04}) the function $p$ will have the form%
\begin{equation*}
p=e^{-\sigma _{1}x_{1}+\sigma _{2}x_{2}},
\end{equation*}%
and the corresponding Vekua equation will be%
\begin{equation}
\partial _{\overline{\zeta }}W-\left( \sigma _{2}-i\sigma _{1}\right) 
\overline{W}=0,  \label{gol08}
\end{equation}%
where $\partial _{\overline{\zeta }}=\frac{\partial }{\partial x_{2}}+i\frac{%
\partial }{\partial x_{1}}.$ Following the Theorems and Definitions cited in
the section of Preliminaries, let us consider the domain $\Omega $ as the
unitary circle, fixing the center of the formal powers at the origin of the
plane $\zeta _{0}=0$ (it is remarkable that an equivalent case was
considered in \cite{raulito}, where the authors performed numerical
calculations in order to examine solutions for boundary value problems for
elliptic differential operators).

Since every formal power $Z^{(m)}\left( a_{m},0;\zeta \right) ,$ $\zeta
=x_{2}+ix_{1},$ is $\left( F,G\right) $-pseudoanalytic, it will be enough to
focus our attention into the electrical currents emerging from the pair of
functions $Z^{(m)}\left( 1,0;\zeta \right) $ and $Z^{(m)}\left( i,0;\zeta
\right) ,$ $=0,1,2,...$, because any other electrical current will be
necessarily a linear combination of them (see Theorem 15).

Hence, let us consider the three first pairs of formal powers, corresponding
to the generating pair $F_{0}=e^{-\sigma _{1}x_{1}+\sigma _{2}x_{2}}$ and $%
G_{0}=ie^{\sigma _{1}x_{1}-\sigma _{2}x_{2}}$, in exact form:%
\begin{equation}
Z^{(0)}\left( 1,0;\zeta \right) =e^{-\sigma _{1}x_{1}+\sigma _{2}x_{2}},%
\text{ \ \ }Z^{(0)}\left( i,0;\zeta \right) =ie^{\sigma _{1}x_{1}-\sigma
_{2}x_{2}};  \label{gol05}
\end{equation}%
\begin{eqnarray}
Z^{(1)}\left( 1,0;\zeta \right) &=&\frac{1}{\sigma _{2}}e^{-\sigma
_{1}x_{1}}\sinh \left( \sigma _{2}x_{2}\right) +\frac{i}{\sigma _{1}}%
e^{-\sigma _{2}x_{2}}\sinh \left( \sigma _{1}x_{1}\right) ,  \label{gol06} \\
Z^{(1)}\left( i,0;\zeta \right) &=&-\frac{1}{\sigma _{1}}e^{\sigma
_{2}x_{2}}\sinh \left( \sigma _{1}x_{1}\right) +\frac{i}{\sigma _{2}}%
e^{\sigma _{1}x_{1}}\sinh \left( \sigma _{2}x_{2}\right) ;  \notag
\end{eqnarray}%
and%
\begin{gather}
Z^{(2)}\left( 1,0;\zeta \right) =\left( \frac{x_{1}}{2\sigma _{1}}+\frac{%
x_{2}}{2\sigma _{2}}\right) e^{-\sigma _{1}x_{1}+\sigma _{2}x_{2}}-
\label{gol07} \\
-\frac{1}{2\sigma _{1}^{2}}e^{\sigma _{2}x_{2}}\sinh \left( \sigma
_{1}x_{1}\right) -\frac{1}{2\sigma _{2}^{2}}e^{-\sigma _{1}x_{1}}\sinh
\left( \sigma _{2}x_{2}\right) +  \notag \\
+\frac{i}{2\sigma _{1}\sigma _{2}}\left( e^{\sigma _{1}x_{1}}\sinh \left(
\sigma _{2}x_{2}\right) -e^{-\sigma _{2}x_{2}}\sinh \left( \sigma
_{1}x_{1}\right) \right) +  \notag \\
+\frac{i}{2\sigma _{1}x_{1}-2\sigma _{2}x_{2}}\left( \frac{x_{1}}{\sigma _{2}%
}-\frac{x_{2}}{\sigma _{1}}\right) \sinh \left( \sigma _{1}x_{1}-\sigma
_{2}x_{2}\right) ,  \notag \\
Z^{(2)}\left( i,0;\zeta \right) =-i\left( \frac{x_{1}}{2\sigma _{1}}+\frac{%
x_{2}}{2\sigma _{2}}\right) e^{\sigma _{1}x_{1}-\sigma _{2}x_{2}}+  \notag \\
+\frac{i}{2\sigma _{1}^{2}}e^{-\sigma _{2}x_{2}}\sinh \left( \sigma
_{1}x_{1}\right) +\frac{i}{2\sigma _{2}^{2}}e^{\sigma _{1}x_{1}}\sinh \left(
\sigma _{2}x_{2}\right) +  \notag \\
+\frac{1}{2\sigma _{1}\sigma _{2}}\left( e^{-\sigma _{1}x_{1}}\sinh \left(
\sigma _{2}x_{2}\right) -e^{\sigma _{2}x_{2}}\sinh \left( \sigma
_{1}x_{1}\right) \right) +  \notag \\
+\frac{1}{2\sigma _{1}x_{1}-2\sigma _{2}x_{2}}\left( \frac{x_{1}}{\sigma _{2}%
}-\frac{x_{2}}{\sigma _{1}}\right) \sinh \left( \sigma _{1}x_{1}-\sigma
_{2}x_{2}\right) .  \notag
\end{gather}

Taking into account the notations introduced in (\ref{gol09}), (\ref{gol01}%
), (\ref{gol03}) and (\ref{gol10}), we have that the current density vectors 
$\overrightarrow{j}^{\left( m\right) }$ will be given by%
\begin{equation}
\overrightarrow{j}^{\left( m\right) }=-\sigma \text{grad}u=\left( 
\begin{array}{c}
\sqrt{\sigma }\mathcal{E}_{1}p^{-1}\text{Re}Z^{\left( m\right) } \\ 
\sqrt{\sigma }\mathcal{E}_{2}p\text{Im}Z^{\left( m\right) } \\ 
0%
\end{array}%
\right) .  \label{gol08a}
\end{equation}

\subsection{Electrical current patches within the unitary circle}

In order to pose a qualitative idea of the current patches inside the
unitary circle, it will be convenient to observe first the traces provoked
by the electrical currents flowing through an homogeneous medium, say $%
\sigma =1$. Evidently, this implies $\sigma _{1}=\sigma _{2}=0$, and thus
the Vekua equation (\ref{gol08}) will turn into the well known
Cauchy-Riemann equation%
\begin{equation*}
\partial _{\overline{\zeta }}W_{h}=0,
\end{equation*}%
for which the standard Taylor series describe the general solution%
\begin{equation*}
W_{h}=\sum_{m=0}^{\infty }a_{m}\zeta ^{m}.
\end{equation*}%
By considering (\ref{gol08a}) for $\zeta ^{0},$ $\zeta $ and $\zeta ^{2}$,
with coefficients $1$ and $i$ alternatively, we will obtain%
\begin{gather}
\overrightarrow{j}_{h}^{\left( 0\right) }\left( 1,0\right) =\left( 
\begin{array}{c}
1 \\ 
0 \\ 
0%
\end{array}%
\right) ,~\overrightarrow{j}_{h}^{\left( 0\right) }\left( i,0\right) =\left( 
\begin{array}{c}
0 \\ 
1 \\ 
0%
\end{array}%
\right) ;  \label{aux} \\
\overrightarrow{j}_{h}^{\left( 1\right) }\left( 1,0\right) =\left( 
\begin{array}{c}
x_{2} \\ 
x_{1} \\ 
0%
\end{array}%
\right) ,\text{ }\overrightarrow{j}_{h}^{\left( 1\right) }\left( i,0\right)
=\left( 
\begin{array}{c}
-x_{1} \\ 
x_{2} \\ 
0%
\end{array}%
\right) ;  \notag \\
\overrightarrow{j}_{h}^{\left( 2\right) }\left( 1,0\right) =\left( 
\begin{array}{c}
x_{2}^{2}-x_{1}^{2} \\ 
2x_{1}x_{2} \\ 
0%
\end{array}%
\right) ,\text{ }\overrightarrow{j}_{h}^{\left( 2\right) }\left( i,0\right)
=\left( 
\begin{array}{c}
-2x_{1}x_{2} \\ 
x_{2}^{2}-x_{1}^{2} \\ 
0%
\end{array}%
\right) .  \notag
\end{gather}%
To trace the electrical current patches for $\overrightarrow{j}_{h}^{\left(
0\right) }\left( 1,0\right) $ and $\overrightarrow{j}_{h}^{\left( 0\right)
}\left( i,0\right) $ is a trivial task, since they posses only one spatial
component. Some of the patches corresponding to the rest of vectors are
shown in the following figures.%
\begin{figure}
\centering
\includegraphics[width=0.70\textwidth]{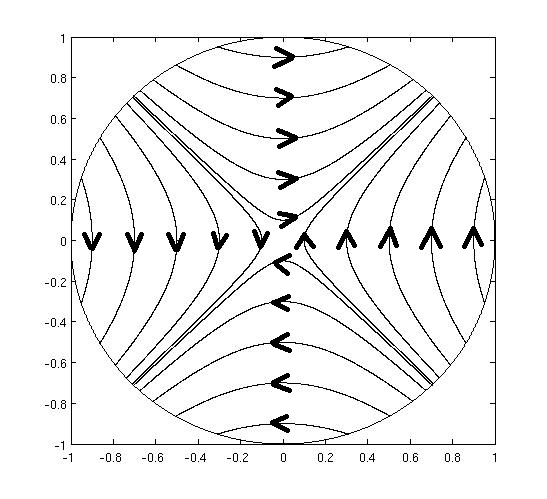}
\caption{$\stackrel{\rightarrow}{j}_{h}^{\left( 1\right) }\left( 1,0\right)$}
\end{figure}

\begin{figure}
\centering
\includegraphics[width=0.70\textwidth]{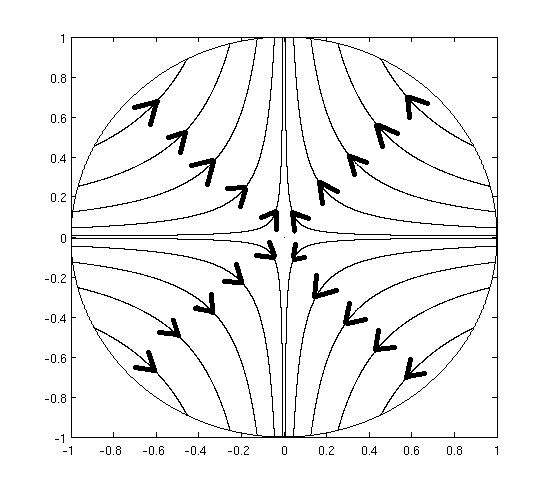}
\caption{$\stackrel{\rightarrow}{j}_{h}^{\left( 1\right) }\left( i,0\right)$}
\end{figure}

\begin{figure}
\centering
\includegraphics[width=0.70\textwidth]{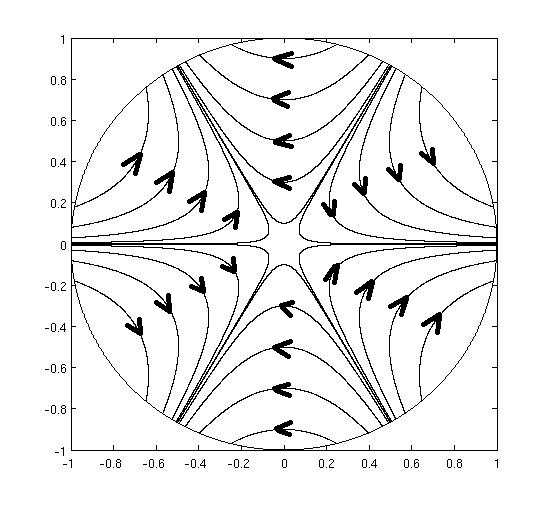}
\caption{$\stackrel{\rightarrow}{j}_{h}^{\left( 2\right) }\left( 1,0\right)$}
\end{figure}

\begin{figure}
\centering
\includegraphics[width=0.70\textwidth]{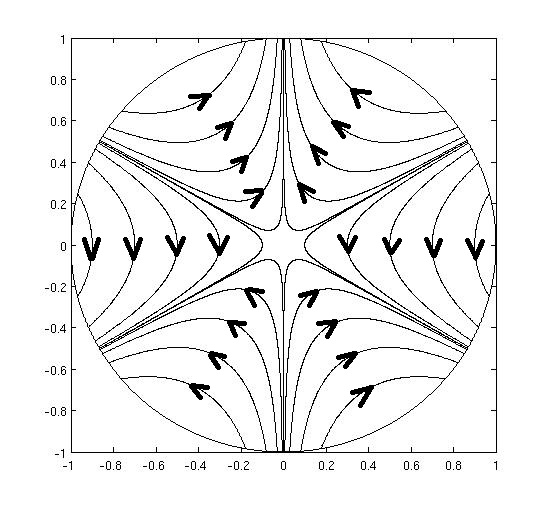}
\caption{$\stackrel{\rightarrow}{j}_{h}^{\left( 2\right) }\left( i,0\right)$}
\end{figure}

In these Figures, as for all corresponding to this section, the arrows point
out the direction of the electrical current flows. Just such nearest to the
center are omitted by virtue of the space.

Let us focus our attention now onto the inhomogeneous case. Just as it
happened before, the traces corresponding to the currents emerging from the
formal powers $Z^{\left( 0\right) }\left( 1,0;\zeta \right) $ and $Z^{\left(
0\right) }\left( i,0;\zeta \right) $ does not reach interesting diagrams.%
\begin{equation}
\overrightarrow{j}^{\left( 0\right) }\left( 1,0;x_{1},x_{2}\right) =\left( 
\begin{array}{c}
e^{2\sigma _{2}x_{2}+2\sigma _{3}x_{3}} \\ 
0 \\ 
0%
\end{array}%
\right) ,\text{ }\overrightarrow{j}^{\left( 0\right) }\left(
i,0;x_{1},x_{2}\right) =\left( 
\begin{array}{c}
0 \\ 
e^{2\sigma _{1}x_{1}+2\sigma _{3}x_{3}} \\ 
0%
\end{array}%
\right) ;  \label{j00}
\end{equation}
More illustrative examples arise when considering the current flows of $%
Z^{\left( 1\right) }\left( 1,0;\zeta \right) $ and $Z^{\left( 1\right)
}\left( i,0;\zeta \right) $ (for simplicity, next Figures are traced by
fixing $x_{3}=0$):%
\begin{gather}
\overrightarrow{j}^{\left( 1\right) }\left( 1,0;x_{1},x_{2}\right) =\left( 
\begin{array}{c}
\frac{1}{\sigma _{2}}e^{\sigma _{2}x_{2}+2\sigma _{3}x_{3}}\sinh \left(
\sigma _{2}x_{2}\right) \\ 
\frac{1}{\sigma _{1}}e^{\sigma _{1}x_{1}+2\sigma _{3}x_{3}}\sinh \left(
\sigma _{1}x_{1}\right) \\ 
0%
\end{array}%
\right) ,  \label{j01} \\
\overrightarrow{j}^{\left( 1\right) }\left( i,0;x_{1},x_{2}\right) =\left( 
\begin{array}{c}
-\frac{1}{\sigma _{1}}e^{\sigma _{1}x_{1}+2\sigma _{2}x_{2}+2\sigma
_{3}x_{3}}\sinh \left( \sigma _{1}x_{1}\right) \\ 
\frac{1}{\sigma _{2}}e^{2\sigma _{1}x_{1}+\sigma _{2}x_{2}+2\sigma
_{3}x_{3}}\sinh \left( \sigma _{2}x_{2}\right) \\ 
0%
\end{array}%
\right) .  \notag
\end{gather}%

\begin{figure}
\centering
\includegraphics[width=0.70\textwidth]{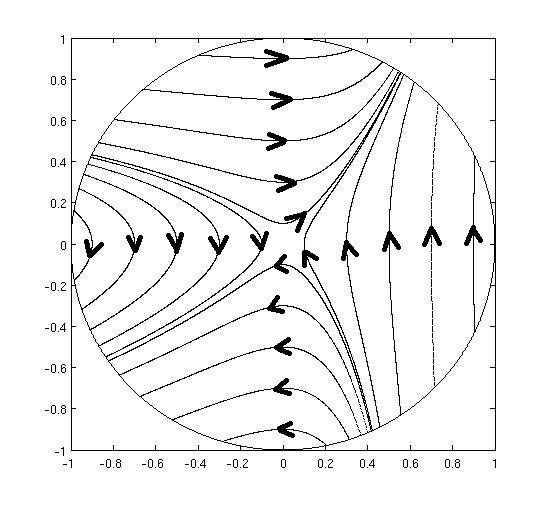}
\caption{$\stackrel{\rightarrow}{j}^{\left( 1\right) }\left( 1,0;x_{1},x_{2}\right)$}
\end{figure}

\begin{figure}
\centering
\includegraphics[width=0.70\textwidth]{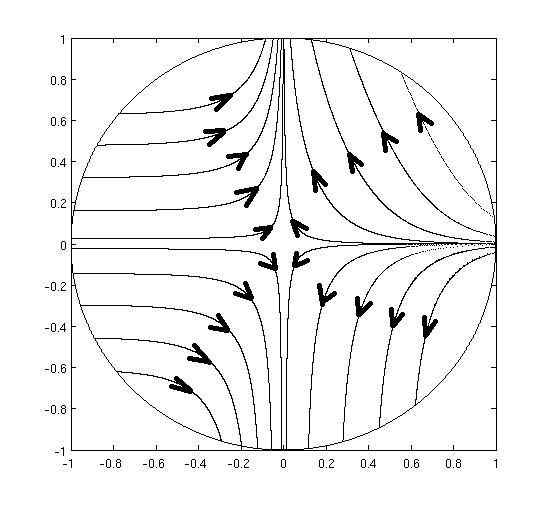}
\caption{$\stackrel{\rightarrow}{j}^{\left( 1\right) }\left( i,0;x_{1},x_{2}\right)$}
\end{figure}

The diagrams are drawn considering $\sigma _{1}=3$ and $\sigma _{2}=1;$ and
by evaluating the current density vectors $\overrightarrow{j}$ into a fixed
point, displacing the mark in the direction of the vector in a distance of $%
0.01\%$ of the Cartesian norm in the subregions with higher conductivity,
and $10\%$ in such with lower.

The Figures 7 and 8 show current patches when considering the remaining formal powers $\overrightarrow{j}^{\left(
2\right) }\left( 1,0;x_{1},x_{2}\right) $ and $\overrightarrow{j}^{\left(
2\right) }\left( i,0;x_{1},x_{2}\right) $.%

\begin{figure}
\centering
\includegraphics[width=0.70\textwidth]{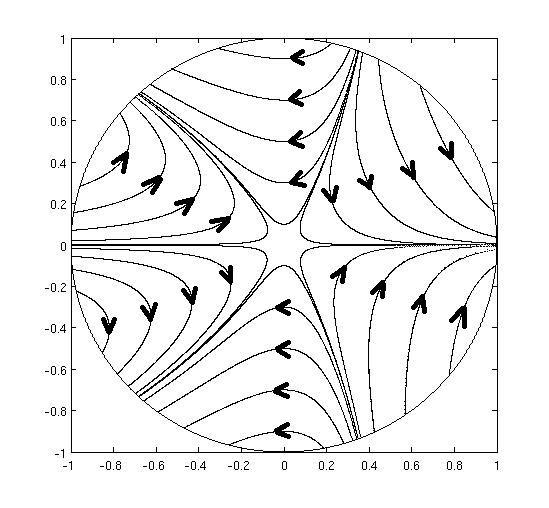}
\caption{$\stackrel{\rightarrow}{j}^{\left( 2\right) }\left( 1,0;x_{1},x_{2}\right)$}
\end{figure}

\begin{figure}
\centering
\includegraphics[width=0.70\textwidth]{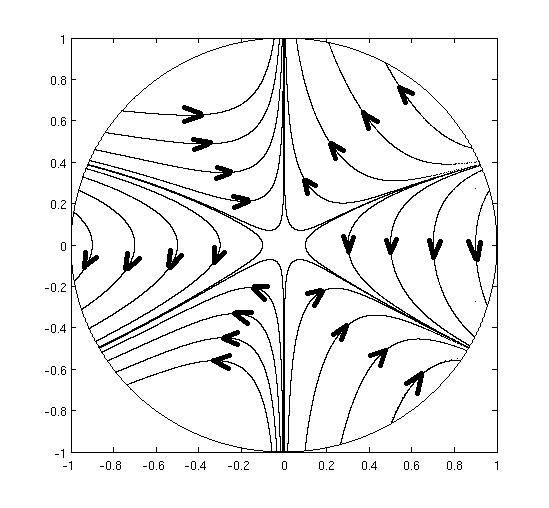}
\caption{$\stackrel{\rightarrow}{j}^{\left( 2\right) }\left( i,0;x_{1},x_{2}\right)$}
\end{figure}

These short previews to the qualitative dynamics of the electrical current
density vectors $\overrightarrow{j}^{\left( m\right) },~m=1,2;$ already
allow us to point out some characteristics that seem to carry out useful
patterns, when comparing them with the currents distributions for a constant
conductivity function.

For instance, Figure 5 keeps the dynamical behavior of the traces found in
Figure 1, presenting the most significant variations into the positive $%
x_{1} $-semiplane. Naturally, the magnitudes of the vector currents in both
figures differ considerably when taking into account the conductivity values
at every point into the domain $\Omega $. Still, a sort of dynamical
relation might be suggested when adopting an appropriate point of view, as
it will be posed in the next section, dedicated to the electric potentials
at the boundary.

It is also remarkable that already watching the traces of the very first
formal powers, we can infer that the most important disparities between the
current patches corresponding to the homogeneous case and such of the
inhomogeneous, will take place on the diagrams belonging to the lower \emph{%
formal degrees.} This is, to such formal powers whose superindex are among
the smallest.

This statement is based onto the location of what could be considered \emph{%
sink points }and \emph{source points, }following the terms employed in
Complex Dynamical Systems (see e.g. \cite{chaos}). Of course, in a strict
sense, not any sink or source point is found inside the unitary circle,
neither will be outside of it, because all traces behave asymptotically.
Nevertheless, the reader could easily infer the location of such regions by
watching the places around the perimeter where the patches seem to get
closer to each other.

The bigger displacement of such regions takes place when appreciating the
diagrams corresponding to the expressions with the smallest formal degrees.
If the reader wishes to verify such behavior, perhaps standard computational
languages for symbolic calculations can be useful in order to obtain in
exact form some formal powers with higher degrees. Other case, it could be
enough to trace again the posed diagrams considering, for example, $\sigma
_{1}=\sigma _{2}=1$.

Thus, from this qualitative appreciation (at least for the inhomogeneous
case just examined), we can observe a \emph{typical} behavior of the formal
powers when compared to the standard powers inside the unitary circle. We
shall remember that this kind of behavior had been already remarked by L.
Bers itself. Indeed, as it was mentioned before, professor Bers gave the
complete proof for the case when $z\rightarrow z_{0}$ \cite{bers}, opening
the opportunity to study the behavior of the formal powers far away from
their center.

\subsection{The electric potential at the boundary}

On the light of the last qualitative overview, a quantitative examination is
in order. Let us pay our attention into one of the most important particular
topics of the Generalized Ohm's Law, when studied into bounded domains: The
behavior of the electric potential $u$ at the boundary.

It is easy to see that for approaching the electric potentials $u^{\left(
m\right) }(x_{1},x_{2})$ corresponding to the current density vectors posed
before, numerical methods are already needed for the cases corresponding to
the formal degree $2$. Let us consider by now some details of the electric
potentials that can be obtained in exact form.

Ignoring the constant parts, the electric potential corresponding to the
current $\overrightarrow{j}^{\left( 0\right) }\left( 1,0;x_{1},x_{2}\right) $
described in (\ref{j00}) is%
\begin{equation*}
u^{\left( 0\right) }\left( 1,0;x_{1},x_{2}\right) =\frac{1}{2\sigma _{1}}%
e^{-2\sigma _{1}x_{1}},
\end{equation*}%
whereas the one for $\overrightarrow{j}^{\left( 0\right) }\left(
i,0;x_{1},x_{2}\right) $ is%
\begin{equation*}
u^{\left( 0\right) }\left( i,0;x_{1},x_{2}\right) =\frac{1}{2\sigma _{2}}%
e^{-2\sigma _{2}x_{2}}.
\end{equation*}%

\begin{figure}
\centering
\includegraphics[width=0.90\textwidth]{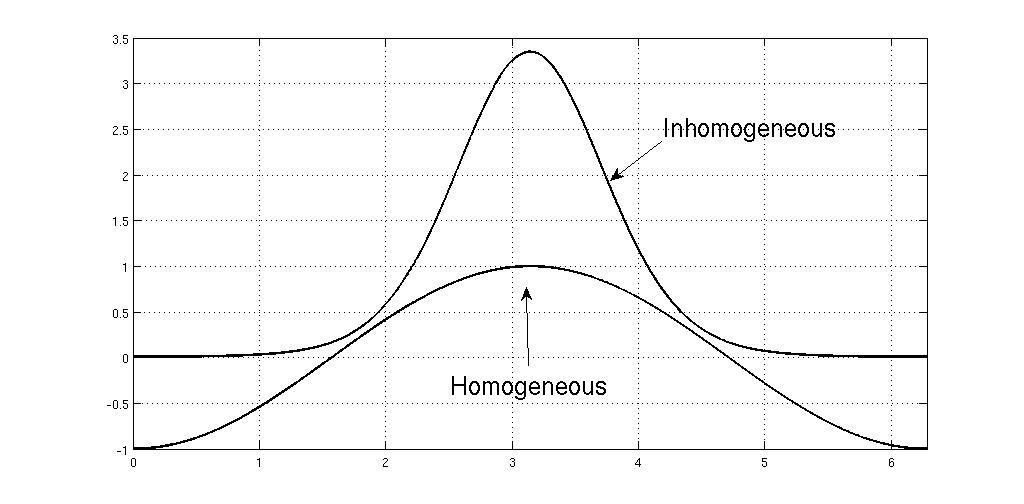}
\caption{$u^{\left( 0\right) }\left( 1,0;x_{1},x_{2}\right)$}
\end{figure}

\begin{figure}
\centering
\includegraphics[width=0.90\textwidth]{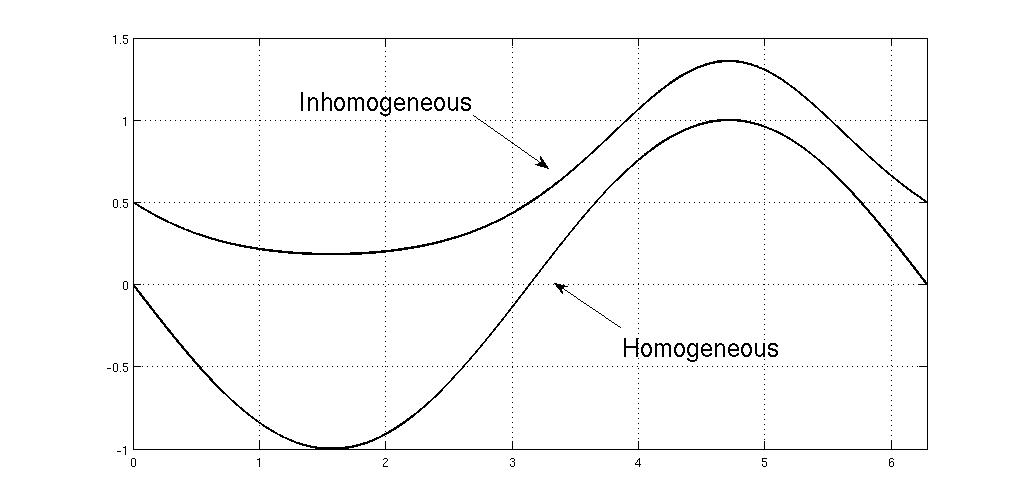}
\caption{$u^{\left( 0\right) }\left( i,0;x_{1},x_{2}\right)$}
\end{figure}

As it is pointed out, the graphics included in Figure 9 correspond to the
electric potential of the inhomogeneous case, titling the Figure with the
abbreviated notation $u^{\left( 0\right) }\left( 1,0\right) ,$ and of the
homogeneous case denoted by $u_{h}^{\left( 0\right) }\left( 1,0\right) $.
The graphics are traced by simply considering the value of the potentials at
the boundary. This is, $x_{1}=\cos \theta $ and $x_{2}=\sin \theta ;$ $%
\theta \in \lbrack 0,2\pi ].$ The same is done for the next Figure, where
the graphics for $u^{\left( 0\right) }\left( i,0\right) $ and $u_{h}^{\left(
0\right) }\left( i,0\right) $ are presented.%

\begin{figure}
\centering
\includegraphics[width=0.90\textwidth]{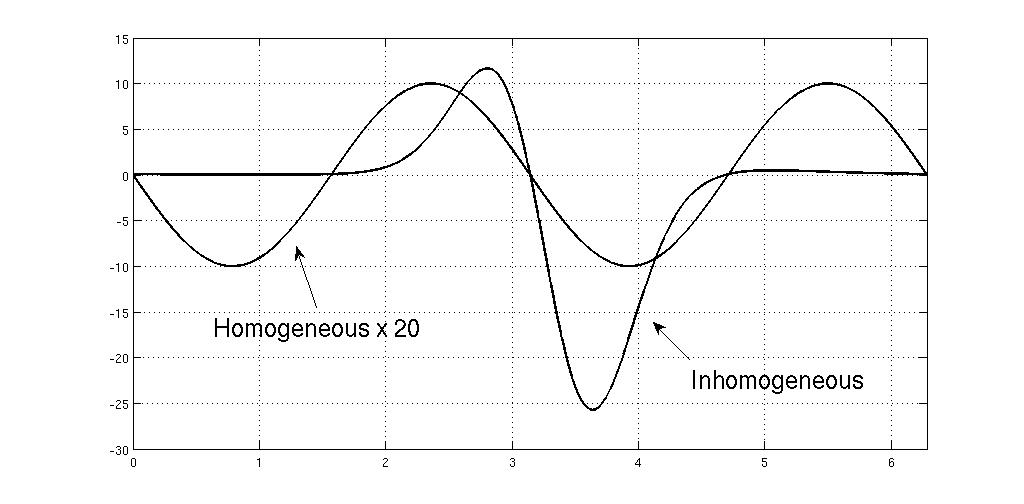}
\caption{$u^{\left( 1\right) }\left( 1,0;x_{1},x_{2}\right)$}
\end{figure}

\begin{figure}
\centering
\includegraphics[width=0.90\textwidth]{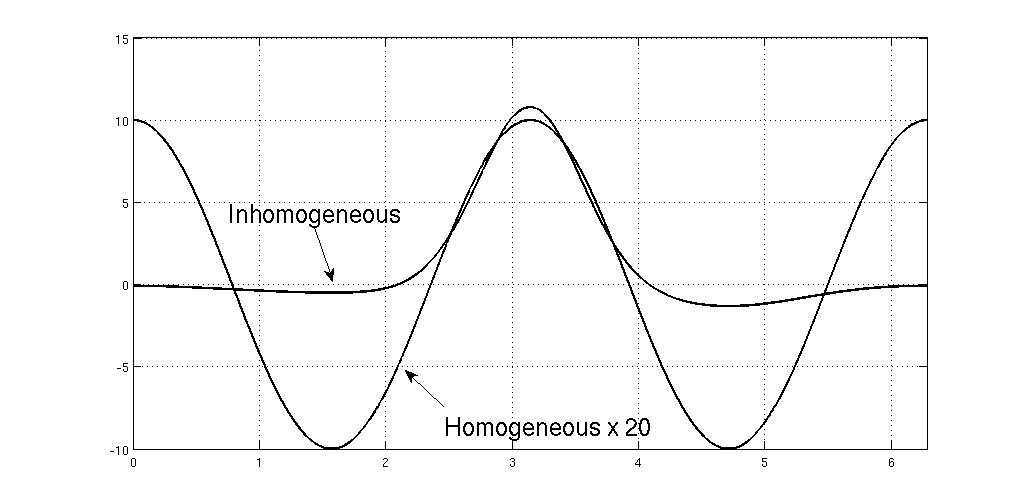}
\caption{$u^{\left( 1\right) }\left( i,0;x_{1},x_{2}\right)$}
\end{figure}

Again, as it was already expected, from a certain point of view it could be
possible to assert that there exist a pattern on the dynamical behavior of
these potentials at the boundary. For these cases, as the following graphics
could suggest, the boundary potentials corresponding to the lower formal
degrees, seem to hold the closest dynamical behavior.

For the current vectors $\overrightarrow{j}^{\left( 1\right) }\left(
1,0;x_{1},x_{2}\right) $ and $\overrightarrow{j}^{\left( 1\right) }\left(
i,0;x_{1},x_{2}\right) $\ of (\ref{j01}), the electric potentials are%
\begin{gather*}
u^{\left( 1\right) }\left( 1,0;x_{1},x_{2}\right) =\frac{1}{4\sigma
_{1}\sigma _{2}}\left( e^{-2\sigma _{1}x_{1}}+e^{-2\sigma
_{2}x_{2}}-e^{-2\sigma _{1}x_{1}-2\sigma _{2}x_{2}}\right) , \\
u^{\left( 1\right) }\left( i,0;x_{1},x_{2}\right) =\frac{x_{1}}{2\sigma _{1}}%
-\frac{x_{2}}{2\sigma _{2}}+\frac{1}{4\sigma _{1}^{2}}e^{-2\sigma _{1}x_{1}}-%
\frac{1}{4\sigma _{2}^{2}}e^{-2\sigma _{2}x_{2}}.
\end{gather*}

The behaviors of the potentials illustrated in Figure 11 and Figure 12,
could still suggest the existence of some dynamical relation, pointing out
that it was already necessary to scale the graphics corresponding to the
homogeneous case by a factor of $20,$ in order to show them together with
such corresponding to the inhomogeneous case.

Other hand, the dynamics performed in this examples can be located nearer of
the behaviors expected when analyzing the electric potentials from a
classical point of view: The variations of the potentials at the boundary
does not hold any simple relation with the variations of the conductivity
inside the domain $\Omega $. Still, by these basic examples, it is possible
to suggest that perhaps from the point of view of the Pseudoanalytic
Function Theory, applied to the study of the Generalized Ohm's Law, the
instability of the electric potential at the boundary, when changes of the
conductivity function inside the domain are taking place, could be
diminished if the changes are considered for every \emph{formal electric
potential} individually.

This could well represent an important contribution for better understanding
inverse problems, as it is the one posed by A. P. Calderon in 1980 \cite%
{calderon}, known in medical imaging as Electrical Impedance Tomography.

\section{A technique for approaching separable-variables conductivity
functions}

A very important problem is located around how to approach a
separable-variable conductivity function once it is given a finite set of
points inside a domain $\Omega $ in the plane, where the electric
conductivity is known (a very natural starting point for many physical
examinations).

This is a critical matter, since all procedures posed in this work are based
upon the idea of possessing an explicit separable-variables conductivity
function. Indeed, it is very difficult to find among the literature any
reference about specific methods for interpolating separable-variables
functions, even for the two-dimensional case. Perhaps the following idea
could serve as a temporally departure point.

Let us suppose we have a finite set of $k_{1}+1$ parallel lines to the $%
x_{1} $-axe, dividing the domain $\Omega $ into $k_{1}$ segments. Suppose we
also have a set of $k_{2}+1$ lines parallel to the $x_{2}$-axe, dividing
again the segments obtained from the first step. The result will be a sort
of grill, with a finite number $k<k_{1}\times k_{1}$ of intersections within
the domain $\Omega $, hence we can introduce a set of points $%
z_{k}=(z_{k}^{\prime },z_{k}^{\prime \prime })$ located precisely at every
intersection inside $\Omega $, where $z_{k}^{\prime }$ represents the $x_{1}$%
-coordinate of $z_{k}$ and $z_{k}^{\prime \prime }$ is the $x_{2}$%
-coordinate. Finally, let us assign a conductivity value $\sigma \left(
z_{k}\right) $ to each point $z_{k}.$

Starting with the segment of line intersecting the $x_{2}$-axe at the
maximum value reached inside $\Omega $, we can interpolate all values $%
\sigma \left( z_{k}\right) $ corresponding to the points $z_{k}$ contained
within that line. Hence we can pose a continuos function of the form%
\begin{equation*}
\alpha_{1}\left( x_{1}\right) =\frac{f_{1}\left( x_{1}\right) }{z_{1}^{\prime
\prime }+K},
\end{equation*}%
where $z_{1}^{\prime \prime }$ is the common $x_{2}$-coordinate of all
points $z_{k}$ belonging to the line segment on which the interpolating
function $f_{1}$ is defined, and $K$ is an arbitrary real constant such that 
$x_{2}+K\neq 0,$ $x_{2}\in \Omega .$

The same can be done for the rest of line segments parallel to $x_{1}$ found
into $\Omega .$

We can then define the following piecewise function $\sigma \left(
x_{1},x_{2}\right) :$%
\begin{equation*}
\sigma \left( x_{1},x_{2}\right) =\left\{ 
\begin{array}{l}
\left( x_{2}+K\right) \alpha_{1}\left( x_{1}\right) :x_{2}\in \left[ \max
x_{2}:x_{2}\in \Omega ,z_{1}^{\prime \prime }-\frac{z_{1}^{\prime \prime
}+z_{2}^{\prime \prime }}{2}\right) ; \\ 
\left( x_{2}+K\right) \alpha_{2}\left( x_{1}\right) :x_{2}\in \left[
z_{1}^{\prime \prime }-\frac{z_{1}^{\prime \prime }+z_{2}^{\prime \prime }}{2%
},z_{2}^{\prime \prime }-\frac{z_{2}^{\prime \prime }+z_{3}^{\prime \prime }%
}{2}\right) ; \\ 
\left( x_{2}+K\right) \alpha_{3}\left( x_{1}\right) :x_{2}\in \left[
z_{2}^{\prime \prime }-\frac{z_{2}^{\prime \prime }+z_{3}^{\prime \prime }}{2%
},z_{3}^{\prime \prime }-\frac{k_{3}^{\prime \prime }+k_{4}^{\prime \prime }%
}{2}\right) ; \\ 
\multicolumn{1}{c}{...} \\ 
\left( x_{2}+K\right) \alpha_{k_{1}}\left( x_{1}\right) :x_{2}\in \left[
z_{k_{1}-1}^{\prime \prime }-\frac{z_{k_{1}-1}^{\prime \prime
}+z_{k_{1}}^{\prime \prime }}{2},\min x_{2}:x_{2}\in \Omega \right] ;%
\end{array}%
\right.
\end{equation*}%
where $z_{k}^{^{\prime \prime }}:k=1,2,...,k_{1}$ are the common $x_{2}$%
-coordinates of every set of $z_{k}$ points belonging to the same line
segment parallel to $x_{1}$ inside $\Omega $.

It is easy to see that the piecewise function $\sigma \left(
x_{1},x_{2}\right) $ defined in the last expression is a separable-variable
function, so it can be employed when numerical calculations are performed
for approaching higher formal powers than those considered in this work.
Notice also that the same idea can be extended for segmentations of a
bounded domain $\Omega $ by means of polar traces.

Moreover, this basic idea could be also useful when considering the
resulting electrical impedance equation when studying the monochromatic
time-dependet case:%
\begin{equation*}
\text{div}\left( \gamma \text{grad}u\right) =0.
\end{equation*}%
Here $\gamma =i\omega \varepsilon +\sigma $ represents the electrical
impedance, $\omega $ is the wave frequency and $\varepsilon $ is a scalar
function denoting the electrical permittivity. The reader can verify that
virtually all mathematical methods explained in the above paragraphs can be
extended for this complex case.

\begin{acknowledgement}
The author would like to acknowledge the support of CONACyT project 106722,
Mexico.
\end{acknowledgement}

\end{document}